\documentclass[10pt,conference]{IEEEtran}
\usepackage{graphicx,amsmath,amssymb,color,bm,setspace,subfigure}
\begin{document}
\newtheorem{numThm}{Theorem}
\newtheorem{numCor}[numThm]{Corollary}
\newtheorem{numLem}[numThm]{Lemma}
\newtheorem{numProp}[numThm]{Proposition}
\newtheorem{numDef}[numThm]{Definition}
\newtheorem{numClaim}[numThm]{Claim}
\newtheorem{numConj}[numThm]{Conjecture}
\newtheorem{numRem}[numThm]{Remark}
\newtheorem{numProb}{Problem}
\renewcommand{\QED}{\QEDopen}
\title{Conditionally Cycle-Free Generalized Tanner Graphs: Theory and Application to High-Rate Serially Concatenated Codes}
\author{\authorblockN{Thomas R. Halford and Keith M. Chugg}
\authorblockA{Communication Sciences Institute\\
Department of Electrical Engineering - Systems\\
University of Southern California\\
Los Angeles, CA 90089-2565\\
Email: \{halford,chugg\}@usc.edu}}
\maketitle
\begin{abstract}
Generalized Tanner graphs have been implicitly studied by a number of authors under the rubric of generalized parity-check matrices.  This work considers the conditioning of binary hidden variables in such models in order to break all cycles and thus derive optimal soft-in soft-out (SISO) decoding algorithms.  Conditionally cycle-free generalized Tanner graphs are shown to imply optimal SISO decoding algorithms for the first order Reed-Muller codes and their duals - the extended Hamming codes - which are substantially less complex than conventional bit-level trellis decoding.  The study of low-complexity optimal SISO decoding algorithms for the family of extended Hamming codes is practically motivated.  Specifically, it is shown that exended Hamming codes offer an attractive alternative to high-rate convolutional codes in terms of both performance and complexity for use in very high-rate, very low-floor, serially concatenated coding schemes.
\end{abstract}
\section{Introduction}
Very high-rate codes are of great interest for a number of practical applications including data storage and high-speed fiber links.  The design of modern error-correcting codes which simultaneously have very low error floors (e.g. $<10^{-10}$ bit error rate) and very high rates  (e.g. $>0.9$) is a particularly challenging problem of practical interest.  Due to the inherent difficulty of simulating the performance of codes in the very low floor region, the design of such codes relies on the principles of uniform interleaver analysis (cf. \cite{BeDiMoPo98b} and the references therein).  To review, on the additive white Gaussian noise (AWGN) channel with binary antipodal signaling, the bit error rate ($P_b$) and codeword error rate ($P_{cw}$) of a large class of modern codes vary asymptotically with block size as
\begin{equation}\label{uniform_int_eqn}
P_b\thicksim N^{\alpha_{\max}}\quad\quad\quad\quad
P_{cw}\thicksim N^{\alpha_{\max}+1}
\end{equation}
where $N$ is the interleaver size and $\alpha_{\max}$ is the maximum exponent of $N$ in an asymptotic union bound approximation. Note that $\alpha_{\max}$ depends on both the specific code construction and constituent codes used.  If the bit (codeword) error rate decays with $N$, then the code is said to exhibit interleaver gain in bit (codeword) error rate.  Designs for codes with low floors require interleaver gain in both bit and codeword error rates thus require $\alpha_{\max}\leq -2$.    

Serially concatenated code (SCC) constructions (i.e. codes composed of an inner code and outer code separated by a random-like interleaver) are well-suited to low-floor design because, provided the inner code is recursive, the maximum exponent of Equation \ref{uniform_int_eqn} is \cite{BeDiMoPo98b}
\begin{equation}
\alpha_{\max}=-\left\lfloor\frac{d_{\min,o}+1}{2}\right\rfloor
\end{equation}
where $d_{\min,o}$ is the minimum distance of the outer code.  Since the rate of an SCC is equal to the product of the rates of its constituent codes, the design of a high-rate, low-floor SCC requires a high-rate outer code satisfying $d_{\min,o}\geq 3$.  However, it is very challenging to find such outer codes  for which there exist low-complexity optimal SISO decoding algorithms.  To this end, Graell i Amat \textit{et al.} introduced a class of high-rate convolutional codes with optimal SISO decoding algorithms of moderate complexity based on their dual codes \cite{GrBeMo02,GrMoBe02}.

An alternative approach to the design of high-rate low-floor codes are the systematic with serially concatenated parity (S-SCP) codes proposed in \cite{ChThDiGrMe05}, of which Jin \textit{et al.}'s generalized repeat accumulate (GRA) codes \cite{JiKhMc00} are an example.  The S-SCP code structure can be viewed as a systematic code with a parity generating concatenated system that resembles an SCC.  It was demonstrated in \cite{ChThDiGrMe05} that S-SCP codes have the same maximum error exponent and design rules as SCCs: codes constructed with a parity generating system composed of a recursive inner parity generator and an outer code satisfying $d_{\min,o}\geq 3$ achieve interleaver gain in both bit and codeword error rates.  In contrast to SCCs, good S-SCPs can be constructed with inner parity generators that have rate greater than $1$ so that the rate of outer code can be \textit{lower} than the overall code rate thus alleviating the aforementioned challenge of finding high-rate, $d_{\min,o}\geq 3$ outer codes with low-complexity SISO decoding algorithms.

The design of good high-rate, low-floor codes has thus been largely solved for the AWGN channel.  However, the S-SCP design philosophy is not directly applicable to the large class of systems which have  recursive channels.  The term \textit{recursive channel} is introduced to describe systems in which the aggregate of the modulation and (possible) precoding with the channel is recursive.  Continuous phase modulations over AWGN and fading channels as well as certain models for precoded magnetic recoding channels (e.g. EPR4 \cite{LiNaKuGe02}) are examples of recursive channels.  

In light of the above discussion, high-rate, low-floor codes for use in systems with recursive channels can be constructed via the serial concatenation of a high-rate code and the channel (where the recursive channel is treated like an inner code).  The outer high-rate code can be a modern (e.g. SCC or S-SCP) code or a classical (e.g. algebraic or convolutional) code.  Classical outer codes are more attractive, however, for application in practical systems.  Specifically, the use of a classical outer code for which there exists a simple non-iterative SISO decoding algorithm offers reductions in decoding complexity, decoding latency, and required memory with respect to a modern, iteratively decoded outer code.  As with the design of SCCs for the AWGN channel, the design of such codes requires a high-rate, $d_{\min}\geq 3$ outer classical code for which there exists a low-complexity optimal SISO decoding algorithm.  

To this end, the present work introduces novel low-complexity SISO decoding algorithms for the family of first-order Reed-Muller codes (and hence their duals, the extended Hamming codes \cite{MaSl78}) in Section \ref{RM_ex_sec} based on variable conditioning in generalized Tanner graphs (GTGs), which are defined in Section \ref{GTG_sec}.  It is shown in Section \ref{TLC_app_sec} that extended Hamming codes offer an attractive alternative to the high-rate convolutional codes studied in \cite{GrBeMo02,GrMoBe02} for use as outer codes in serial concatenation with recursive inner channels.  Concluding remarks are given in Section \ref{conc_sec}.
\section{Conditionally Cycle-Free GTGs}\label{GTG_sec}
\subsection{Generalized Code Extensions}
Let $\mathcal{C}$ be an $[n,k,d]$ binary linear block code and let
\begin{equation}
\mathcal{I}\subseteq\left\{1,2,\ldots,n\right\} 
\end{equation}
be some subset of the coordinate index set of $\mathcal{C}$.  A \textit{generalized extension} of $\mathcal{C}$, $\widetilde{\mathcal{C}}$, is formed by adding a parity-check on the bits corresponding to the index set $\mathcal{I}$ to $\mathcal{C}$ (i.e. a \textit{partial} parity-check).  Note that if $\mathcal{I}=\left\{1,2,\ldots,n\right\}$ then $\widetilde{\mathcal{C}}$ is simply  a classically defined \textit{extended code} \cite{MaSl78}.  The generalized extended code $\widetilde{\mathcal{C}}$ has length $n+1$, dimension $k$ and minimum distance either $d$ or $d+1$ depending on the choice of $\mathcal{I}$.  More generally, a \textit{degree-g generalized extension} of $\mathcal{C}$ is formed by adding $g$ partial parity bits to $\mathcal{C}$.  Note that the $j^{th}$ partial parity bit, $c_{n+j}$, in such an extension forms a partial parity on any of the bits $c_i$ for $i\in\{1,\ldots,n+j-1\}$.
\subsection{Generalized Tanner Graphs}
This work introduces the term \textit{generalized Tanner graph} (GTG) to denote the Tanner graph\footnote{The term Tanner graph has been used to describe different classes of graphical models by different authors.  In this work, the term Tanner graph is used in its most restricted sense.  That is to say, there is a bijection between the set of Tanner graphs for an $[n,k,d]$ binary linear code $\mathcal{C}$ and the set of $n-k\times n$ parity check matrices for that code.} \cite{Ta81} corresponding to the parity-check matrix of a generalized code extension.  Specifically, let $\widetilde{\mathcal{C}}$ be a degree-$g$ generalized extension of the $[n,k,d]$ binary linear block code $\mathcal{C}$ and let $\widetilde{H}=\left[h_{ij}\right]$ be an $\left(n-k+g\right)\times\left(n+g\right)$ parity-check matrix for $\widetilde{\mathcal{C}}$.  The GTG associated with $\widetilde{H}$ is the bipartite graph $G_{\widetilde{H}}=(\mathcal{U}\cup\mathcal{W},\mathcal{E})$ with disjoint vertex classes:
\begin{equation}
\mathcal{U}=\left\{u_j\right\}_{j=1}^{n+g}\quad\mbox{and}\quad\mathcal{W}=\left\{w_i\right\}_{i=1}^{n-k+g} 
\end{equation}
corresponding to the columns and rows of $\widetilde{H}$, respectively.  An edge connects $u_j$ and $w_i$ in $G_{\widetilde{H}}$ if and only if $h_{ij}=1$.  Note that a number of authors have previously considered such graphical representations of binary codes under the rubric of \textit{generalized parity-check matrices} (cf. \cite{YeChFo02}).

\begin{figure}[htbp]
\centering
\includegraphics[width=3.00in]{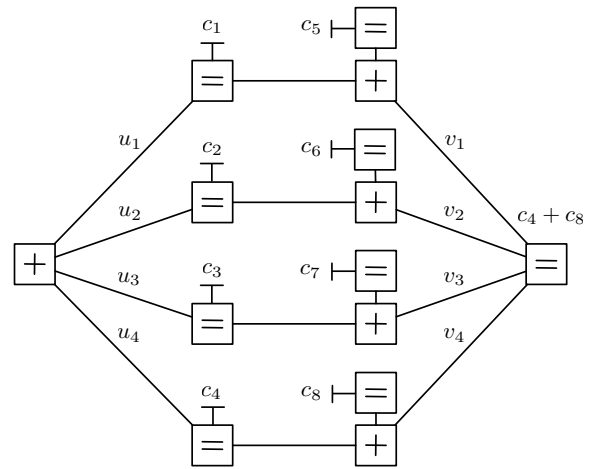}
\caption{Generalized Tanner graph for the $[8,4,4]$ first-order Reed-Muller code implied by a squaring construction.  Repetition and single parity-check constraints are labeled with `=' and `+' symbols, respectively.}
\label{eh_plotkin_fig}
\end{figure}

As an example, consider the $[8,4,4]$ first-order Reed-Muller code $\mathcal{C}_{RM(1,3)}$, any codeword of which can be decomposed via the squaring construction \cite{MaSl78} as
\begin{equation}
\begin{split}
\mathbf{c}&=\left(c_1,c_2,c_3,c_4,c_5,c_6,c_7,c_8\right)\\
&=\left(u_1,u_2,u_3,u_4,u_1+v_1,u_2+v_2,u_3+v_3,u_4+v_4\right)\\
&=\left(\mathbf{u},\mathbf{u}+\mathbf{v}\right)
\end{split}
\end{equation}
where $\mathbf{u}$ is drawn from the $[4,3,2]$ single parity-check code and $\mathbf{v}$ is drawn from the $[4,1,4]$ repetition code.  A degree-$1$ generalized extension $\widetilde{\mathcal{C}}_{RM(1,3)}$ can be formed by adding the partial parity bit $c_4+c_8=v_4$ to $\mathcal{C}_{RM(1,3)}$.  A parity-check matrix for $\widetilde{\mathcal{C}}_{RM(1,3)}$ is
\begin{equation}\label{rm13_pc_mx3}
\widetilde{H}_{RM(1,3)}=\left[\begin{array}{cccccccccc}
1&1&1&1&0&0&0&0&0\\
1&0&0&0&1&0&0&0&1\\
0&1&0&0&0&1&0&0&1\\
0&0&1&0&0&0&1&0&1\\
0&0&0&1&0&0&0&1&1
\end{array}\right].
\end{equation}
Figure \ref{eh_plotkin_fig} illustrates the GTG corresponding to $\widetilde{H}_{RM(1,3)}$.
\subsection{Conditionally Cycle-Free Generalized Tanner Graphs}\label{CCF_GTG_sec}
Hidden variable conditioning in graphical models is well-understood in the context of tail-biting trellises (cf. \cite{AjHoMc98}).  If a tail-biting trellis contains a hidden (state) variable, $V$, with alphabet size $2^m$ then optimal SISO decoding can be performed on this trellis by decoding on $2^m$ cycle-free conditional trellises (one per possible value of $V$) and then appropriately marginalizing over the results of the decoding rounds.  This work applies hidden variable conditioning to generalized Tanner graphs in order to obtain cycle-free graphical models and thus optimal SISO decoding algorithms.

\begin{figure}[ht]
\begin{center}
\includegraphics[width=2.5in]{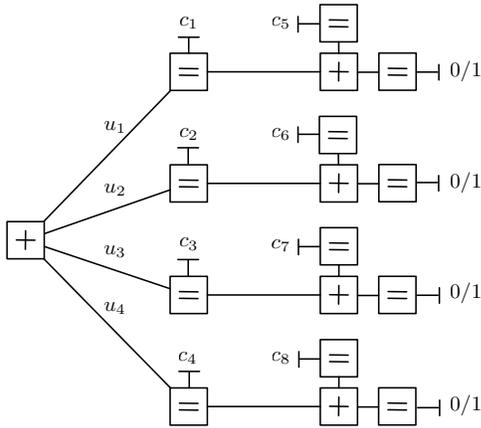}
\caption{Conditionally cycle-free generalized Tanner graph for the $[8,4,4]$ first-order Reed-Muller code.}
\label{eh_plotkin_fixed_fig}
\end{center}
\end{figure}

As an example, consider again the GTG for $\mathcal{C}_{RM(1,3)}$ illustrated in Figure \ref{eh_plotkin_fig} and suppose that the partial parity variable $c_4+c_8$ is not considered unknown but fixed to be either a $0$ or a $1$.  This \textit{conditional} GTG is illustrated in Figure \ref{eh_plotkin_fixed_fig}.  Note that the conditional variable $c_4+c_8$ is now treated as a visible, deterministic variable as indicated by the half-edges labeled by `$0/1$' in Figure \ref{eh_plotkin_fixed_fig}.  
The GTG illustrated in Figure \ref{eh_plotkin_fixed_fig} is cycle-free; optimal SISO decoding of a code with such a model can be accomplished by decoding twice on the conditionally cycle-free model (once with $c_4+c_8=0$ and once with $c_4+c_8=1$) and then appropriately marginalizing over the results of the two decoding rounds \cite{HeCh01}.

More generally, a \textit{degree-f conditionally cycle-free} generalized Tanner graph contains $f$ variables which can be fixed to remove all cycles.  Optimal SISO decoding can be achieved by decoding $2^f$ times on a degree-$f$ conditionally cycle-free GTG (and appropriately marginalizing over the results).  
\section{Conditionally Cycle-Free GTGs for First-Order Reed-Muller Codes}\label{RM_ex_sec}
Let $\mathcal{C}_{RM(r,m)}$ denote the Reed-Muller code with parameters
\begin{equation}
n=2^m,\quad k=\sum_{j=0}^r\binom{m}{j},\quad d=2^{m-r}.
\end{equation}
\begin{numThm}\label{RM_acyclic_thm}
There exists a degree-$(m-2)$ conditionally cycle-free generalized Tanner graph for $\mathcal{C}_{RM(1,m)}$, $m\geq 3$.

\begin{proof}
By induction on $m$.  The $m=3$ case was shown by construction in Section \ref{GTG_sec}.  Suppose that there exists a degree-$(m-3)$ conditionally cycle-free graphical model for $\mathcal{C}_{RM(1,m-1)}$.  There exists a squaring construction for $\mathcal{C}_{RM(1,m)}$ such that if $\mathbf{c}\in\mathcal{C}_{RM(1,m)}$ then
\begin{equation}
\mathbf{c}=\left(\mathbf{u},\mathbf{u}+\mathbf{v}\right)
\end{equation}
where $\mathbf{u}\in\mathcal{C}_{RM(1,m-1)}$ and $\mathbf{v}\in\mathcal{C}_{RM(0,m-1)}$ \cite{MaSl78}.  A GTG for this code is illustrated in Figure \ref{general_RM_plotkin_fig}.  The code vertex labeled $\mathcal{C}_{RM(1,m-1)}$ corresponds to a GTG for that code.
\begin{figure}[htbp]
\begin{center}
\includegraphics[width=3.25in]{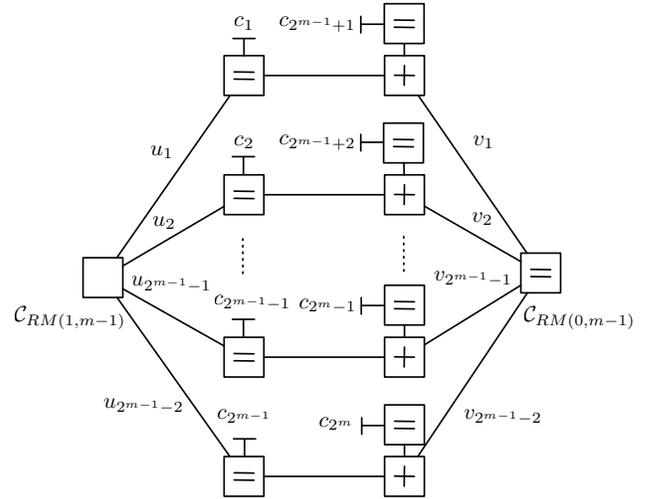}
\caption{Generalized Tanner graph for the $[2^m,m+1,2^{m-1}]$ Reed-Muller code implied by a squaring construction.}
\label{general_RM_plotkin_fig}
\end{center}
\end{figure}

By the induction hypothesis, a degree-$(m-2)$ conditionally cycle-free GTG is formed by conditioning on the repetition constraint $\mathcal{C}_{RM(0,m-1)}$ and on $m-3$ variables in the model corresponding to $\mathcal{C}_{RM(1,m-1)}$.
\end{proof}
\end{numThm}
\vspace{3pt}

It is well-known that a symbol-wise optimal SISO decoding algorithm for a code $\mathcal{C}$ implies a symbol-wise optimal SISO decoding algorithm for its dual $\mathcal{C}^\bot$ \cite{HaRu76,MoBe01}.  The conditionally cycle-free GTGs for first-order Reed-Muller codes described above thus imply optimal SISO decoding algorithms for the family of Reed-Muller codes with parameters $[2^m,2^m-m-1,4]$: the extended Hamming codes.  

Let $A_m$ and $C_m$ denote the number of addition and comparison\footnote{Note that min-sum or min$^\star$-sum processing is assumed throughout this work so that $C_m$ counts the number of min or min$^\star$ operations.} operations required by the optimal SISO decoding algorithm implied by the conditionally cycle-free GTG for $\mathcal{C}_{RM(1,m)}$ described above.  Counting operations in the binary graphical model illustrated in Figure \ref{general_RM_plotkin_fig} yields the following recursion for $A_m$ and $C_m$:
\begin{equation}
A_m=3\cdot 2^m+2A_{m-1},\quad
C_m=2^m+2C_{m-1}.
\end{equation}
Note that these formulae include the comparisons required for marginalizing over the results of the individual decoding rounds.  It is clear that the complexity of SISO decoding using the proposed conditionally acyclic binary graphical models thus grows as $\mathcal{O}\left(n\log n\right)$.   Ashikhim and Litsyn proposed a MAP decoding algorithm for the family of $\mathcal{C}_{RM(1,m)}$ codes with similar asymptotic complexity based on fast Hadamard transforms \cite{AsLi04}.  Note that the complexity of bit-level trellis decoding of this code family grows as $\mathcal{O}\left(n^2\right)$. 

In order to specifically demonstrate that conditionally cycle-free generalized Tanner graphs (CCF-GTGs) can imply substantially less complex optimal SISO decoding algorithms than bit-level trellises, the number of addition and comparison operations for each SISO decoding algorithm is tabulated in Table \ref{RM_siso_table} for a number of first-order Reed-Muller codes.  The trellis complexity was evaluated by considering the total number of addition and comparison operations required to perform the full SISO BCJR algorithm \cite{BaCoJeRa74} (rather that soft-in hard-out Viterbi decoding) on bit-level trellises with minimal state complexities as given by Lu and Huang in \cite{LuHu95}.    
\begin{table}[h]
\begin{center}
\begin{tabular}{c||c|c||c|c}
&\multicolumn{2}{c||}{\small{Trellis}}&\multicolumn{2}{c}{\small{CCF-GTG}}\\
\cline{2-5}
&\small{Add}&\small{Compare}&\small{Add}&\small{Compare}\\
\hline\hline
$\mathcal{C}_{RM(1,3)}$&\small{176}&\small{50}&\small{72}&\small{32}\\
$\mathcal{C}_{RM(1,4)}$&\small{1008}&\small{426}&\small{192}&\small{80}\\
$\mathcal{C}_{RM(1,5)}$&\small{4720}&\small{2202}&\small{480}&\small{192}\\
$\mathcal{C}_{RM(1,6)}$&\small{20336}&\small{9850}&\small{1152}&\small{448}\\
$\mathcal{C}_{RM(1,7)}$&\small{84336}&\small{41530}&\small{2688}&\small{1024}\\
$\mathcal{C}_{RM(1,8)}$&\small{343408}&\small{170426}&\small{6144}&\small{2304}\\
\end{tabular}
\caption{Complexity of trellis decoding vs. the proposed optimal SISO decoding algorithm.}\label{RM_siso_table}
\end{center}
\end{table}
\vspace{-30pt}
\section{Application: High-Rate SCCs}\label{TLC_app_sec}
In \cite{GrMoBe02}, Graell i Amat \textit{et al.} studied a serially concatenated coding scheme consisting of a high-rate convolutional outer code and a recursive rate-$1$ inner code corresponding to a simplified discrete-time model of a precoded EPR4 magnetic recording channel.  Specifically, the recursive channel model comprises a $1/1\oplus D^2$ precoder followed by a digital recording channel subject to intersymbol interference (ISI) with partial response polynomial $1+D-D^2-D^3$ followed finally by an AWGN channel.  Note that the precoder and ISI can be jointly decoded on an $8$-state trellis.  

In this section, the performance of this scheme is compared to one which replaces the high-rate convolutional codes with extended Hamming codes.  Specifically, $4$ convolutional outer codes with input block size $4000$ bits and respective rates $8/9$, $9/10$, $11/12$, and $16/17$ are compared to four algebraic outer codes composed of the following mixtures of extended Hamming codes:
\begin{itemize}
\item The mixture of $3$ $\mathcal{C}_{RM(3,5)}$ and $69$ $\mathcal{C}_{RM(4,6)}$ codewords resulting in a code with input block size $4011$ bits and rate $4011/4512=0.8890\approx 8/9$.
\item The mixture of $56$ $\mathcal{C}_{RM(4,6)}$ and $7$ $\mathcal{C}_{RM(5,7)}$ codewords resulting in a code with input block size $4032$ bits and rate $4032/4480=9/10$.
\item The mixture of $30$ $\mathcal{C}_{RM(4,6)}$ and $19$ $\mathcal{C}_{RM(5,7)}$ codewords resulting in a code with input block size $3990$ bits and rate $3990/4352=0.9168\approx 11/12$.
\item The mixture of $2$ $\mathcal{C}_{RM(4,6)}$, $26$ $\mathcal{C}_{RM(5,7)}$, and $3$ $\mathcal{C}_{RM(6,8)}$ codewords resulting in a code with input block size $3975$ bits and rate $3975/4224=0.9411\approx 16/17$.
\end{itemize}
As reported in \cite{GrMoBe02}, the SCCs using convolutional codes utilize s-random interleavers \cite{DiPo95}.  The codes using mixed extended Hamming outer codes utilize high sum-spread pseudo-random interleavers that were generated using the real-relaxation optimization method described in \cite{Cr00}.

\begin{figure}[htbp]
\begin{center}
\includegraphics[width=3.00in]{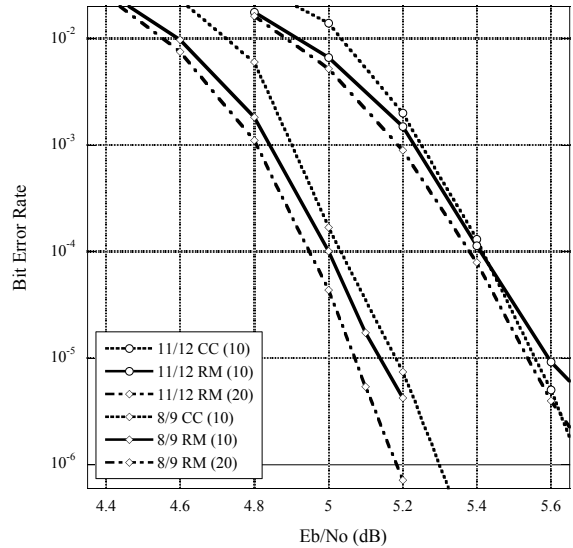}
\caption{Bit error rate performance of the respective rate $8/9$ and $11/12$ serially concatenated codes.}
\label{results_2_fig}
\end{center}
\end{figure}

\begin{figure}[htbp]
\begin{center}
\includegraphics[width=3.00in]{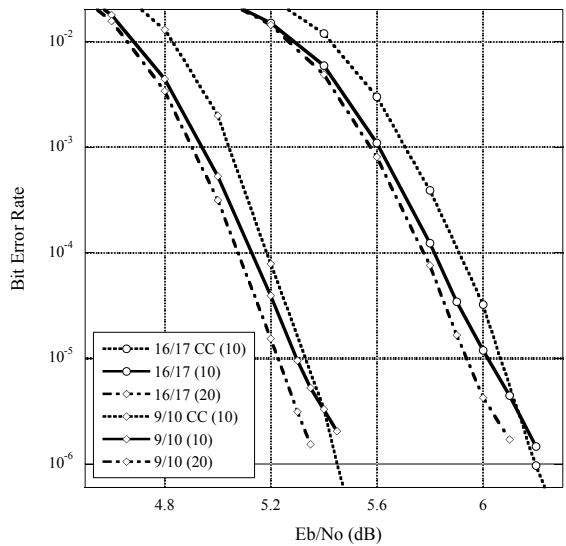}
\caption{Bit error rate performance of the respective rate $9/10$ and $16/17$ serially concatenated codes.}
\label{results_1_fig}
\end{center}
\end{figure}

Figure \ref{results_2_fig} compares the performance of the respective rate $8/9$ and $11/12$ codes while Figure \ref{results_1_fig} compares the performance of the respective rate $10/11$ and $16/17$ codes.  Note that the performance of the serially concatenated codes with convolutional outer codes is reported for $10$ decoding iterations while the performance of the codes with mixed extended Hamming outer codes is reported for $10$ and $20$ decoding iterations.  All curves correspond to min$^\star$-sum processing (or its dual-domain equivalent \cite{MoBe01}). 

Observe in Figures \ref{results_2_fig} and \ref{results_1_fig} that the codes constructed with mixed extended Hamming outer codes compare favorably in terms of performance to those with convolutional outer codes of similar rates and input block sizes.  The mixed extended Hamming codes compare favorably to their high-rate convolutional code counterparts in terms of complexity also.   Table \ref{RM_siso_table_2} tabulates the average number of add and comparison operations per input bit, per decoding iteration, required for optimal SISO decoding (using the CCF-GTG algorithm) for each of the mixed extended Hamming codes.  The rate $k/k+1$ convolutional outer-codes in \cite{GrMoBe02} were decoded on the $16$- (for $k=8,9,10$) and $32$- (for $k=16$) state trellises corresponding to their respective rate $1/1+k$ duals.  Optimal SISO decoding on a rate $1/1+k$, $16$- ($32$-) state trellis requires at least $96$ ($192$) additions and $64$ ($128$) comparisons per input bit, per decoding iteration.  Thus, if one assumes that an addition and comparison operation have roughly the same complexity\footnote{This assumption is clearly reasonable for min-sum processing. For min$^\star$-star processing, this assumption is also reasonable provided a table-lookup is used for the correction term in the min$^\star$ operator.}, then the complexity of the proposed mixed extended Hamming code SISO decoding algorithms are approximately 5 to 10 times less than that of the respective high-rate convolutional code decoding algorithms proposed in \cite{GrMoBe02}.
\begin{table}[h]
\begin{center}
\begin{tabular}{c||c|c}
Rate &Add / Bit&Compare / Bit\\
\hline\hline
\small{8/9}&\small{20.2}&\small{7.9}\\
\small{9/10}&\small{20.7}&\small{8.0}\\
\small{11/12}&\small{21.5}&\small{8.2}\\
\small{16/17}&\small{22.8}&\small{8.7}
\end{tabular}
\caption{Average number of operations per decoding iteration, for optimal SISO decoding of the mixed extended Hamming codes.}\label{RM_siso_table_2}
\end{center}
\end{table}
\vspace{-24pt}
\section{Conclusion}\label{conc_sec}
Motivated by the search for optimal SISO decoding algorithms with complexity less than that of trellis decoding, this work studied conditionally cycle-free generalized Tanner graphs.  It was shown that such models exist for the family of first-order Reed-Muller codes and that these models imply low-complexity SISO decoding algorithms for these codes and their duals: the extended Hamming codes.  It was shown that the proposed low-complexity SISO decoding algorithms for the family of extended Hamming codes are particularly useful in the context of high-rate serially concatenated codes.  

A number of interesting directions for future work are motivated by the present work.  It would be interesting to search for more codes for which conditionally cycle-free generalized Tanner graphs offer complexity savings over trellis decoding.  It would also be interesting to study the performance of the sub-optimal, iterative, SISO decoding algorithms implied by GTGs in which only a \textit{subset} of hidden variables required to break all cycles are conditioned.  Specifically, a study of codes with known algebraic decompositions may lead to novel low-complexity SISO decoding algorithms.  There are also a number of interesting avenues of study concerning the application of the optimal SISO decoder for extended Hamming codes.  Specifically, the algorithm presented in this work has a natural tree structure which can lead to particularly efficient implementations in hardware \cite{BeCh01}.
%\bibliographystyle{../IEEETran}
%\bibliography{../IEEEabrv,../chugg_strings,../chugg_group_2006,../kmc_books,../kmc_conf_submitted,../kmc_conf,../kmc_journal,../kmc_journal_submitted,../halford_to_merge}

\end{document}